\documentclass[prl,twoside,preprintnumbers,superscriptaddress,twocolumn,nofootinbib,nobibnotes]{revtex4}
\usepackage{amsmath,graphicx,slashed,multirow,color,ulem}
\usepackage{graphicx,epsfig,amssymb,bm,slashed,wasysym,multirow}
\usepackage[utf8]{inputenc}
\def \beq{\begin{equation}}
\def \eeq{\end{equation}}
\def \beqa{\begin{eqnarray}}
\def \eeqa{\end{eqnarray}}

\def\bea{\begin{eqnarray}}
\def\eea{\end{eqnarray}}

\def\gsim{\mathrel{\rlap{\lower4pt\hbox{\hskip1pt$\sim$}}
    \raise1pt\hbox{$>$}}}         
\def\lsim{\mathrel{\rlap{\lower4pt\hbox{\hskip1pt$\sim$}}
    \raise1pt\hbox{$<$}}}         

\preprint{NIKHEF 2016-050}

\begin{document}
\title{Precision determination of the small-$x$ gluon from  charm production at LHCb}

\author{Rhorry Gauld}
\email{rgauld@phys.ethz.ch}
\affiliation{Institute for Theoretical Physics, ETH, CH-8093 Zurich, Switzerland}
\affiliation{Institute for Particle Physics Phenomenology, University of Durham, DH1 3LE Durham, United Kingdom}

\author{Juan Rojo}
\email{j.rojo@vu.nl}
\affiliation{Department of Physics and Astronomy, VU University Amsterdam, De Boelelaan 1081, NL-1081, HV Amsterdam, The Netherlands}
\affiliation{Nikhef, Science Park 105, NL-1098 XG Amsterdam, The Netherlands}

\date{\today}

\begin{abstract}
  The small-$x$ gluon in global fits of parton distributions
  is affected by large uncertainties from the lack of direct experimental constraints.
  In this work we provide a precision determination of the small-$x$ gluon
  from the exploitation of forward charm production data provided by LHCb 
  for three different centre-of-mass (CoM) energies: 5~TeV, 7~TeV and 13~TeV.
  The LHCb measurements are included in the PDF fit by means of
  normalized distributions and cross-section ratios between data taken
  at different CoM values, $R_{13/7}$ and $R_{13/5}$.
  We demonstrate that forward charm production leads to a reduction
  of the PDF uncertainties of the  gluon down to $x\simeq 10^{-6}$
  by up to an order of magnitude, with implications
  for high-energy colliders, cosmic ray physics and neutrino astronomy.
\end{abstract}

\maketitle


The determination of the internal structure of the proton,
as encoded by the non-perturbative parton distribution functions
(PDFs)~\cite{Forte:2013wc,Butterworth:2015oua,Rojo:2015acz},
has far-reaching implications for
many areas in nuclear, particle and astroparticle physics.
A topic that has recently attracted substantial interest is the
determination of the gluon PDF at small-$x$, which
is of direct relevance for the modelling of soft QCD at the LHC~\cite{Skands:2014pea},
neutrino astronomy~\cite{CooperSarkar:2011pa,Garzelli:2015psa,Gauld:2015kvh,Bhattacharya:2016jce} 
and cosmic ray physics~\cite{d'Enterria:2011kw}, as well as for future 
lepton-proton~\cite{AbelleiraFernandez:2012cc} and proton-proton
higher-energy colliders~\cite{Mangano:2016jyj}.
Constraints on the gluon PDF from deep-inelastic scattering
(DIS) inclusive and charm structure functions at
HERA~\cite{Abramowicz:1900rp,Abramowicz:2015mha}
are limited to $x \gsim 3\cdot 10^{-5}$ in the perturbative
region, and consequently for smaller values of $x$ there are large
uncertainties from the lack of direct experimental information.

Last year, it was realized~\cite{Gauld:2015yia,Zenaiev:2015rfa,Cacciari:2015fta}
that a way forward was provided by considering
inclusive $D$ meson production in $pp$ collisions at the LHC, 
for which the LHCb experiment had already provided data 
at 7~TeV~\cite{Aaij:2013mga}.
The inclusive charm cross-section at the LHC is dominated 
by heavy quark pair production, in turn driven
by the gluon-gluon luminosity, 
and the forward LHCb kinematics 
allow a coverage of the small-$x$ region
that can reach as low as $x\simeq 10^{-6}$.
While the direct inclusion of absolute $D$ meson cross-sections
into a PDF fit is unfeasible due to the large theory
uncertainties that affect the NLO calculation,
it has been demonstrated~\cite{Zenaiev:2015rfa,Gauld:2015yia}
that by using tailored normalized distributions it is possible
to exploit the LHCb measurements
to achieve a significantly improved
description of the small-$x$ gluon.
A complementary approach, suggested in~\cite{Cacciari:2015fta},
would be to include $D$ meson data into PDF fits with the use of ratios
of cross-sections between different center-of-mass (CoM)
energies, which benefit from various uncertainty
cancellations~\cite{Mangano:2012mh}.

More recently, the LHCb collaboration has presented the analogous $D$ meson
cross-section measurements at $\sqrt{s}=5$
and 13~TeV~\cite{Aaij:2016jht,Aaij:2015bpa}, together with the corresponding
ratios $R_{13/7}$ and $R_{13/5}$.\footnote{The LHCb 5 and 13~TeV data
  considered here corresponds to the update provided in May 2017.}
In this letter, we quantify the
impact of the LHCb $D$ meson data at different CoM energies
on the small-$x$ gluon from the NNPDF3.0 global analysis~\cite{Ball:2014uwa}.
These data are included both
in terms of normalized cross-sections as well as by means of the
cross-section ratio measurements.
Our strategy leads to a precision determination
of the small-$x$ gluon, substantially improving previous results,
and highlighting the consistency of the LHCb measurements at the three
CoM energies.
We illustrate the implications of our results for 
ultra high-energy (UHE) neutrino-nucleus cross-sections $\sigma_{\nu N}(E_{\nu})$,
and the longitudinal structure function $F_L(x,Q^2)$ at future lepton-proton colliders.



The LHCb $D$ meson production data
is presented double differentially in transverse momentum ($p_T^D$) and
rapidity ($y^D$) for a number of final states,
$D^0, D^+, D_s^+$ and $D^{*+}$, which also
contain the contribution from charge-conjugate states.
To include these measurements into the global PDF fit, we define
two observables:
\begin{align}
  \label{eq:Obs}
N_X^{ij} &=  \frac{d^2\sigma({\rm X~TeV})}{dy_i^D d (p_T^D)_j} \bigg{/} \frac{d^2\sigma({\rm X~TeV})}{dy_{\rm ref}^D d (p_T^D)_j} \, , \\[1mm]
R_{13/X}^{ij} &= \frac{d^2\sigma({\rm 13~TeV})}{dy_i^D d (p_T^D)_j} \bigg{/} \frac{d^2\sigma({\rm X~TeV})}{dy_{\rm i}^D d (p_T^D)_j} \, ,
\end{align}
which benefit from the partial cancellation of the
residual scale dependence from missing higher-orders, while
retaining sensitivity to the gluon since different
regions of $x$ are probed in the numerator and denominator
of these observables.
The ratio measurements, $R_{13/7}$ and $R_{13/5}$,
are available for $y^D \in [2.0,4.5]$ in five bins and for 
$p_T^D \in [0,8]~{\rm~GeV}$ in eight bins.
The 5~TeV and 13~TeV absolute cross
section measurements extend to  higher $p_T^D$  values, however
these additional points are excluded from the fit since
they might be affected by large logarithmic contributions~\cite{Cacciari:1993mq}.
In this way, data in a fixed kinematic region is included either through
cross-section ratios or normalized cross-sections.
The reference rapidity bin in the normalized distributions $N_X^{ij}$ in Eq.~(\ref{eq:Obs})
is chosen to be $y^D_{\rm ref} \in [3.0,3.5]$, as in~\cite{Zenaiev:2015rfa},
since we have verified that this choice maximizes
the cancellation of scale uncertainties for the considered data.
We restrict our analysis to the $\{ D^0, D^+, D_s^+ \}$ final states, ignoring
that of $D^{+*}$ which has an overlapping contribution with that of $D^0$ and $D^+$.

The theoretical predictions for $D$ meson production are computed
at {\sc\small NLO$+$PS} accuracy using
{\sc\small POWHEG}~\cite{Nason:2004rx,Frixione:2007vw,Alioli:2010xd}
to match the fixed-order calculation~\cite{Frixione:2007nw} to the 
{\sc\small Pythia8} shower~\cite{Sjostrand:2007gs,Sjostrand:2014zea} 
with the {\sc\small Monash 2013} tune~\cite{Skands:2014pea}.
The {\sc\small POWHEG} results have previously
been shown to be consistent~\cite{Cacciari:2012ny,Gauld:2015yia}
with both the NLO+PS (a){\sc\small MC@NLO}~\cite{Frixione:2002ik,Alwall:2014hca} method
and the semi-analytic {\sc\small FONLL} 
calculation~\cite{Cacciari:1998it,Cacciari:2001td}.
The NNPDF3.0 NLO set of parton distributions with $\alpha_s(m_Z) = 0.118$, $N_f=5$
and $N_{\rm rep}=1000$ replicas has been used, interfaced
with {\sc\small LHAPDF6}~\cite{Buckley:2014ana}.
The internal {\sc\small POWHEG} routines have been modified to extract
$\alpha_s$ from {\sc\small LHAPDF6}, and the compensation
terms~\cite{Cacciari:1998it} to consistently match the $N_f=5$ PDFs with the fixed-order
$N_f=3$ calculation~\cite{Frixione:2007nw} are included.
The central value for the charm quark pole mass is taken to
be $m_c = 1.5~{\rm~GeV}$, following the 
{\sc\small HXSWG} recommendation~\cite{deFlorian:2016spz}, 
and the renormalization and factorization scales
are set equal to the heavy quark transverse 
mass in the Born configuration,
$\mu=\mu_R=\mu_F=\sqrt{m_c^2+p_T^2}$.

Other settings of the theory calculation,
such as the values for fragmentation
fractions, are the same as those in~\cite{Gauld:2015yia}.
We have verified that the choice of {\sc\small Pythia8} tune 
(comparing {\sc\small Monash 2013} with {\sc\small 4C} or 
{\sc\small A14}) as well as the 
modelling of charm fragmentation (using for {\it e.g.} a
Peterson function with $\epsilon_D = 0.05$ and varying
$\epsilon_D$ by a factor 2) on the observables of Eq.~(\ref{eq:Obs}) leads 
in all cases to variations that are negligible as compared to PDF uncertainties.



The impact of the LHCb $D$ meson data on the NNPDF3.0 small-$x$ gluon
can be quantified using the Bayesian reweighting
technique~\cite{Ball:2010gb,Ball:2011gg}.
We have studied separately the impact of
the three data sets of normalized distributions, $N_5$,
$N_7$ and $N_{13}$ and the two cross-section ratios,
$R_{13/5}$ and $R_{13/7}$, as well specific combinations
of these, always avoiding double counting.
The experimental bin-by-bin correlation matrices
are included for the cross-section ratios $R_{13/X}$, while
for the normalized cross-section data the (cross-section level) 
bin-by-bin correlations, which are only available for 
$N_5$ and $N_{13}$, are not included.

We find that NLO theory describes successfully both
the cross-section ratios $R_{13/7}$ and $R_{13/5}$
as well as the normalized cross section data at
all three CoM energies.
To illustrate this agreement,
we compute the $\chi^2/N_{\rm dat}$ for each of the five datasets,
for different combinations of data used as input in the PDF fit.
These results are summarized in Table~\ref{tab:chi2}, where the
data that has been included in each case are highlighted
in boldface, and the number in brackets indicates $N_{\rm dat}$
for each data set.
For example, the first row corresponds
to the baseline PDF set, the second row indicates the
resultant $\chi^2/N_{\rm dat}$ for each data set after the
$N_{5}$ data has been added to NNPDF3.0, and so on.

\renewcommand*{\arraystretch}{1.2}
\begin{table}[h!]
	\centering
	\begin{tabular}{ c c c c c@{}}
      	\hline
	 $N_{5}(84)$ & $N_{7}(79)$ & $N_{13}(126)$ & $R_{13/5}(107)$ & $R_{13/7}(102)$ \\ \hline \hline	 
	 $1.97$ & $1.21$ & $2.36$ & $1.36$ & $0.80$ \\ \hline
        $\mathbf{0.86}$&  $0.72$ 			&  $1.14$ 			& $1.35$			& $0.81$	\\ 
        $1.31$ 		&  $\mathbf{0.91}$ 	&  $1.58$ 			& $1.36$ 			& $0.82$	\\ 
        $0.74$ 		&  $0.66$ 			&  $\mathbf{1.01}$	& $1.38$ 			& $0.80$	\\ 
        $1.08$ 		&  $0.81$ 			&  $1.27$ 			& $\mathbf{1.29}$	& $0.80$	\\
        $1.53$ 		&  $0.99$ 			&  $1.73$			& $1.30$			& $\mathbf{0.81}$	\\ \hline
        $\mathbf{1.07}$	& $0.81$		& $1.34$ 				& $1.35$ 			& $\mathbf{0.81}$	\\ 
        $0.82$ 			& $\mathbf{0.70}$ &  $1.07$		& $\mathbf{1.35}$	& $0.81$	\\ 
        $\mathbf{0.84}$ 	& $\mathbf{0.71}$&  $\mathbf{1.10}$	& $1.36$			& $0.81$	\\ 
        \hline
	\end{tabular}
        \caption{The $\chi^2/{N_{\rm dat}}$
          for the LHCb $D$ meson measurements considered, $N_5$, $N_7$, $N_{13}$, $R_{13/7}$ and $R_{13/5}$, for various
          combinations of input to the PDF fit (highlighted
          in boldface).
        } \label{tab:chi2}
\end{table}

\begin{figure}[h]
  \begin{center}
    \makebox{\includegraphics[width=1.0\columnwidth]{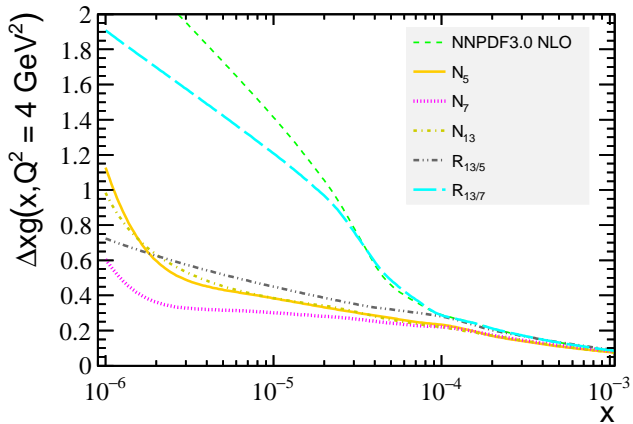}}
  \end{center}
  \vspace{-0.7cm}
  \caption{The 1-$\sigma$ relative PDF uncertainties for the small-$x$
    gluon at $Q^2=4$ GeV$^2$ in NNPDF3.0 and in the subsequent fits
    when the LHCb charm data
    are included one at a time.
  }
  \label{fig:gPDF_unc_comb}
\end{figure}

We find that the normalized distributions, 
$N_5$, $N_7$ and $N_{13}$, as well as the ratio $R_{13/5}$, have a similar
substantial pull on the gluon, both for central values and for the
reduction of the PDF uncertainty.
It is found that the $R_{13/7}$ ratio data has only a minor impact
on the central value and resultant uncertainty of the 
small-$x$ gluon. This can in part be understood due to the fact
that this data is less precise in comparison to the $R_{13/5}$ data,
and additionally less sensitive to the rate of change of the gluon PDF
at low-$x$.
We find it reassuring that including each of the available LHCb 
data sets to NNPDF3.0, one at a time, 
improves the description of all other data sets.
In Fig.~\ref{fig:gPDF_unc_comb} we
show the 1-$\sigma$ relative PDF uncertainties for the 
gluon at $Q^2=4$ GeV$^2$ in NNPDF3.0 and in the subsequent fits
when the various LHCb $D$ meson data sets
are included.

\begin{figure}[h]
  \begin{center}
    \makebox{\includegraphics[width=1.0\columnwidth]{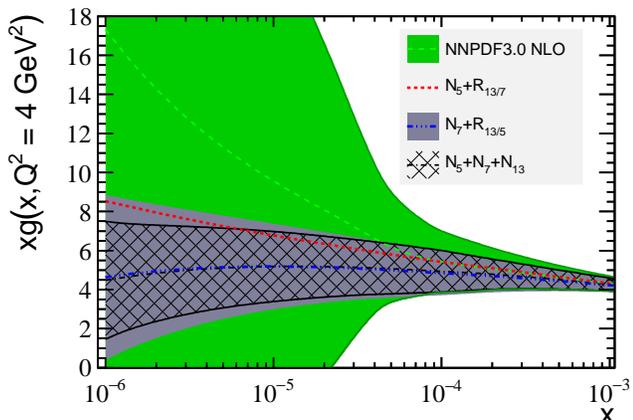}}
  \end{center}
  \vspace{-0.8cm}
  \caption{The NLO gluon in NNPDF3.0 and for
    various combinations of LHCb data included, at $Q^2=4$ GeV$^2$.
  }
  \label{fig:Gluon}
\end{figure}

\begin{figure}[h]
  \begin{center}
    \makebox{\includegraphics[width=1.0\columnwidth]{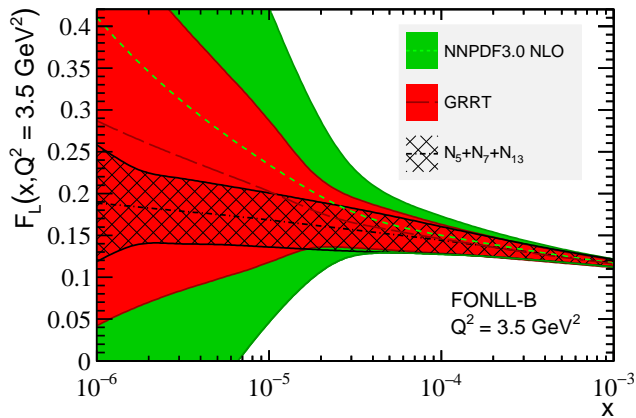}}
  \end{center}
  \vspace{-0.7cm}
  \caption{The structure function $F_L(x,Q^2)$
    at $Q^2=3.5$ GeV$^2$,
    comparing the NNPDF3.0 predictions both
    with the results of this work and with the
    GRRT calculation.
  }
  \label{fig:FL}
\end{figure}

In the following we show results for two representative combinations
of the LHCb measurements, namely $N_7+R_{13/5}$ and $N_5+N_7+N_{13}$.
In Fig.~\ref{fig:Gluon} we compare the small-$x$ gluon in NNPDF3.0 with the 
resultant gluon in these two cases, as well as
the central value from the $N^5+R^{13/7}$ fit.
The central value of the small-$x$ gluon is consistent for all
three combinations, down to $x\simeq 10^{-6}$,
and, as expected from Fig.~\ref{fig:gPDF_unc_comb},
we observe a dramatic reduction of the 1-$\sigma$ PDF uncertainties.
We have verified that these updated results are consistent
with our original study~\cite{Gauld:2015yia}  (GRRT),
yet significantly more precise, as demonstrated in Fig.~\ref{fig:FL}.

\begin{figure}[h]
  \begin{center}
    \makebox{\includegraphics[width=1.0\columnwidth]{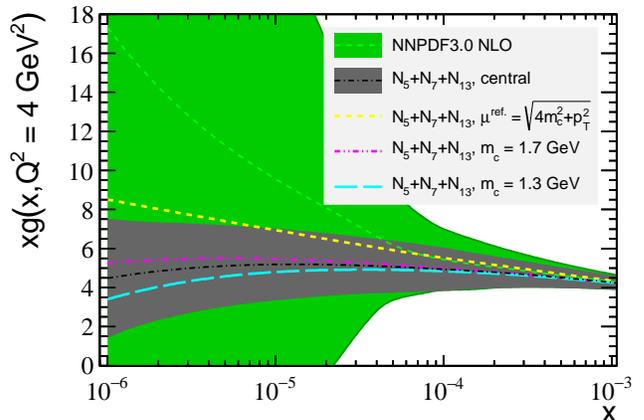}}
  \end{center}
  \vspace{-0.8cm}
  \caption{Dependence of the small-$x$ gluon
    from the $N_5+N_7+N_{13}$ fits for
     variations in the input theory settings.}
  \label{fig:Uncertainty}
\end{figure}

\begin{figure}[h]
  \begin{center}
    \makebox{\includegraphics[width=1.0\columnwidth]{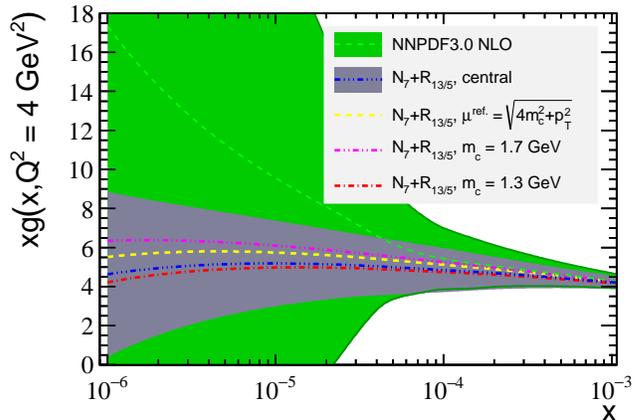}}
  \end{center}
  \vspace{-0.8cm}
  \caption{Same as Fig.~\ref{fig:Uncertainty} for
the $N^7+R^{13/5}$ fits.
  }
  \label{fig:Uncertainty2}
\end{figure}

Given the sizeable theory errors that affect charm production,
it is important to assess the robustness of our results with
respect to the scale variations of the NLO calculation as well as
with the value of $m_c$.
We thus have quantified how the resultant gluon
are affected by theory variations, including: $\mu=\sqrt{4m_c^2+p_T^2}$, 
as well as charm mass variations of $\Delta m_c=0.2$~GeV.
An additional check of applying a minimum $p_T$ requirement
of $2$~GeV to the cross section data was also performed, which
had little impact and is not shown.
The resultant central values of the gluon are shown
in Figs.~\ref{fig:Uncertainty} and~\ref{fig:Uncertainty2},
compared with NNPDF3.0 and with the 1-$\sigma$ PDF
uncertainty band from the
$N_5+N_7+N_{13}$ and $N_7+R_{13/5}$ fits, respectively.

We find that our results are reasonably stable
upon these variations of the input theory settings,
in particular for the $N^7+R^{13/5}$ fits, highlighting that
the cancellation of theory errors is more effective
for the cross-section ratios than for the normalized
distributions.
Even for the most constraining combination,
corresponding to $N_5+N_7+N_{13}$, all theory variations
are contained within the 95\% confidence level interval
of the PDF uncertainty.
This study demonstrates that the sizeable reduction of the small-$x$
gluon PDF errors is robust with respect to theoretical
uncertainties.
A further reduction of the scale dependence could only
be achieved by the full NNLO calculation, so far available 
only for $t\bar{t}$ production~\cite{Czakon:2016ckf}.

Our precision determination of the small-$x$ gluon has
important phenomenological implications, which we 
choose to illustrate with two representative examples: 
the longitudinal structure function $F_L$ at a
future high-energy lepton-proton collider,
and the UHE neutrino-nucleus cross-section.
First of all, we have computed $F_L(x,Q^2)$  for $Q^2=3.5$ GeV$^2$
using {\sc\small APFEL}~\cite{Bertone:2013vaa}
in the FONLL-B general mass scheme~\cite{Forte:2010ta}.
The proposed Large Hadron electron Collider (LHeC) would be able
to measure to measure $F_L$ down to
$x\gsim 10^{-6}$ with
few percent precision
for $Q^2 \gsim 2$ GeV$^2$~\cite{AbelleiraFernandez:2012cc},
hence providing a unique probe
of BFKL resummations
and non-linear QCD dynamics~\cite{Rojo:2009ut}.
In Fig.~\ref{fig:FL} we compare $F_L$ computed with
NNPDF3.0 and with the results of this work, as well
as with the original GRRT calculation.
We observe that the PDF uncertainties on $F_L$ at $x\simeq 10^{-6}$
are now reduced by around a factor of 5 with respect to the baseline, 
and that $F_L$ itself is always positive 
for the $x$ range accessible at the LHeC.

Next, we have computed the UHE charged-current (CC)
neutrino-nucleus
cross-section as a function of the incoming neutrino energy
$E_{\nu}$, using a stand-alone
code based on {\sc\small APFEL} for the calculation of the
NLO structure functions.
At the highest values of $E_{\nu}$ that might be accessible
at neutrino telescopes such as IceCube~\cite{Aartsen:2016ngq}
and KM3NET~\cite{Adrian-Martinez:2016fdl}, the
neutrino-nucleus interactions probes the quark sea PDFs
at $Q^2 \simeq M_W^2$ and down to $x\simeq 10^{-8}$, a region
where the quark distributions are driven by the small-$x$ gluon 
by means of DGLAP evolution effects~\cite{das}.

\begin{figure}[h]
  \begin{center}
    \makebox{\includegraphics[width=1.0\columnwidth]{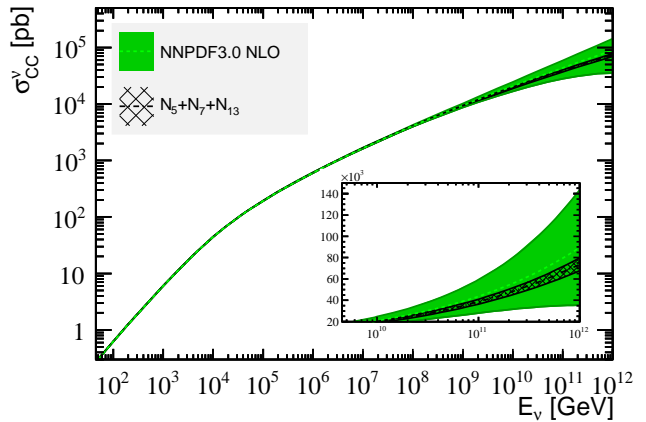}}
  \end{center}
  \vspace{-0.7cm}
  \caption{The NLO charged-current neutrino-nucleus
    cross-section as a function of the neutrino energy $E_{\nu}$,
    computed with NNPDF3.0 and with the results of this work.
  }
  \label{fig:neutrino}
\end{figure}

In Fig.~\ref{fig:neutrino} we compare the CC UHE neutrino-nucleus
cross-section from NNPDF3.0 with the results of this work.
As in the case of $F_L$, we find a sizeable reduction of the
PDF uncertainties, which are by far the dominant theory uncertainty for
this process at high $E_{\nu}$.
This way, NLO QCD provides a prediction accurate to $\lsim 10\%$ up
to $E_{\nu}\simeq 10^{12}$ GeV, a region
where a rather different
behaviour are found in scenarios with non-linear QCD
evolution effects~\cite{Albacete:2015zra}.
Our results for the UHE cross-section are 
more precise than existing
calculations~\cite{Connolly:2011vc},
based on PDF fits where the only constraints on the small-$x$ gluon
come from the inclusive and charm HERA data, and therefore provide
a clean handle to disentangle possible beyond the Standard
Model effects in this process~\cite{Anchordoqui:2013dnh}.

To summarize, in this work we have presented a precision
determination of the small-$x$ gluon down to $x\simeq 10^{-6}$ from LHCb charm production
in the forward region at $\sqrt{s}=$5, 7 and 13~TeV.
We have shown that the LHCb data provided at the three CoM energies leads
to consistent constraints on the small-$x$ gluon.
It is found that the combination of normalized cross section data available at the 
three CoM energies (namely $N_5+N_7+N_{13}$) leads to the strongest constraints 
on the low-$x$ gluon PDF. While this result is shown to be reasonably stable
upon theory variation (see Fig.~\ref{fig:Uncertainty}), a future analysis at NNLO would also be desirable.
Consistent results are also found when a combination
of normalized cross-section and cross-section ratio data are included, the 
most constraining in this case being $N_7+R_{13/5}$, and
found to be robust with respect to theory variations.
Our analysis provides a strong motivation to include 
the LHCb charm production data in the next generation 
of global PDF fits.

We have illustrated how the improved small-$x$ gluon leads to significantly
reduced theory uncertainties for $F_L$ at future high-energy lepton-proton
colliders and for the UHE neutrino-nucleus interactions.
We have however only scratched the surface of the phenomenological
implications of our work. It is important to explore these implications 
further to inform other applications, such as the modelling of semi-hard
QCD processes at the LHC in Monte Carlo event generators and for
calculations of cosmic ray production.
Moreover,
it would be interesting to compare our determination of the small-$x$ gluon
with those that could be achieved from other processes with
similar kinematical coverage, such as exclusive production~\cite{Jones:2016ldq} or
forward photon production~\cite{d'Enterria:2012yj,Peitzmann:2016gkt}.

The results of this work are  available upon request
in the form of {\sc\small LHAPDF6} grids~\cite{Buckley:2014ana}.

{\it Acknowledgements.}
We thank L.~Rottoli and V.~Bertone for the calculations
of the UHE neutrino cross-sections and for assistance with {\tt APFEL}.
We are also grateful to Dominik M\"uller and Alex Pearce for
information with regards to the LHCb data.
We acknowledge the
support provided by the GridPP Collaboration.
The work of J.~R. is partially supported by an ERC Starting Grant ``PDF4BSM".
The work of R.~G. is supported by the ERC Advanced Grant MC@NNLO (340983).

\newpage
{\it Comments regarding update of LHCb data.}
After completing our original analysis of the LHCb data
in October 2016, it was pointed by one of us that there 
were a number of inconsistencies present in the heavy
flavour production data provided by LHCb~\cite{Gauld:2017omh}.
The charm cross-section measurements at 5 and 13~TeV~\cite{Aaij:2016jht,Aaij:2015bpa} 
were subsequently updated, and the current analysis includes
these corrected data.

In our original analysis we observed tension between 
the NLO theory predictions and the LHCb data
for $N_{5}$ and $N_{13}$ observables in particular kinematic
regions. In particular, this discrepancy was observed for rapidity 
bins which were farthest from the reference bin $y_{\rm ref}^D$ which
normalises the cross section data, and at the time this was 
attributed to a limitation of the NLO theory predictions.
Consequently, we applied a kinematic cut to restrict our analysis 
to exclude these regions. This tension is not observed in the update 
of the LHCb data, and no such kinematic restriction is applied in our analysis.
It is likely we were observing the effects of the incorrect
rapidity dependent efficiency correction which affected the LHCb data.

\bibliography{HVQ}

\begin{thebibliography}{51}
\expandafter\ifx\csname natexlab\endcsname\relax\def\natexlab#1{#1}\fi
\expandafter\ifx\csname bibnamefont\endcsname\relax
  \def\bibnamefont#1{#1}\fi
\expandafter\ifx\csname bibfnamefont\endcsname\relax
  \def\bibfnamefont#1{#1}\fi
\expandafter\ifx\csname citenamefont\endcsname\relax
  \def\citenamefont#1{#1}\fi
\expandafter\ifx\csname url\endcsname\relax
  \def\url#1{\texttt{#1}}\fi
\expandafter\ifx\csname urlprefix\endcsname\relax\def\urlprefix{URL }\fi
\providecommand{\bibinfo}[2]{#2}
\providecommand{\eprint}[2][]{\url{#2}}

\bibitem[{\citenamefont{Forte and Watt}(2013)}]{Forte:2013wc}
\bibinfo{author}{\bibfnamefont{S.}~\bibnamefont{Forte}} \bibnamefont{and}
  \bibinfo{author}{\bibfnamefont{G.}~\bibnamefont{Watt}},
  \bibinfo{journal}{Ann.Rev.Nucl.Part.Sci.} \textbf{\bibinfo{volume}{63}},
  \bibinfo{pages}{291} (\bibinfo{year}{2013}), \eprint{1301.6754}.

\bibitem[{\citenamefont{Butterworth et~al.}(2016)}]{Butterworth:2015oua}
\bibinfo{author}{\bibfnamefont{J.}~\bibnamefont{Butterworth}}
  \bibnamefont{et~al.}, \bibinfo{journal}{J. Phys.}
  \textbf{\bibinfo{volume}{G43}}, \bibinfo{pages}{023001}
  (\bibinfo{year}{2016}), \eprint{1510.03865}.

\bibitem[{\citenamefont{Rojo et~al.}(2015)}]{Rojo:2015acz}
\bibinfo{author}{\bibfnamefont{J.}~\bibnamefont{Rojo}} \bibnamefont{et~al.},
  \bibinfo{journal}{J. Phys.} \textbf{\bibinfo{volume}{G42}},
  \bibinfo{pages}{103103} (\bibinfo{year}{2015}), \eprint{1507.00556}.

\bibitem[{\citenamefont{Skands et~al.}(2014)\citenamefont{Skands, Carrazza, and
  Rojo}}]{Skands:2014pea}
\bibinfo{author}{\bibfnamefont{P.}~\bibnamefont{Skands}},
  \bibinfo{author}{\bibfnamefont{S.}~\bibnamefont{Carrazza}}, \bibnamefont{and}
  \bibinfo{author}{\bibfnamefont{J.}~\bibnamefont{Rojo}},
  \bibinfo{journal}{European Physical Journal} \textbf{\bibinfo{volume}{74}},
  \bibinfo{pages}{3024} (\bibinfo{year}{2014}), \eprint{1404.5630}.

\bibitem[{\citenamefont{Cooper-Sarkar et~al.}(2011)\citenamefont{Cooper-Sarkar,
  Mertsch, and Sarkar}}]{CooperSarkar:2011pa}
\bibinfo{author}{\bibfnamefont{A.}~\bibnamefont{Cooper-Sarkar}},
  \bibinfo{author}{\bibfnamefont{P.}~\bibnamefont{Mertsch}}, \bibnamefont{and}
  \bibinfo{author}{\bibfnamefont{S.}~\bibnamefont{Sarkar}},
  \bibinfo{journal}{JHEP} \textbf{\bibinfo{volume}{08}}, \bibinfo{pages}{042}
  (\bibinfo{year}{2011}), \eprint{1106.3723}.

\bibitem[{\citenamefont{Garzelli et~al.}(2015)\citenamefont{Garzelli, Moch, and
  Sigl}}]{Garzelli:2015psa}
\bibinfo{author}{\bibfnamefont{M.~V.} \bibnamefont{Garzelli}},
  \bibinfo{author}{\bibfnamefont{S.}~\bibnamefont{Moch}}, \bibnamefont{and}
  \bibinfo{author}{\bibfnamefont{G.}~\bibnamefont{Sigl}},
  \bibinfo{journal}{JHEP} \textbf{\bibinfo{volume}{10}}, \bibinfo{pages}{115}
  (\bibinfo{year}{2015}), \eprint{1507.01570}.

\bibitem[{\citenamefont{Gauld et~al.}(2016)\citenamefont{Gauld, Rojo, Rottoli,
  Sarkar, and Talbert}}]{Gauld:2015kvh}
\bibinfo{author}{\bibfnamefont{R.}~\bibnamefont{Gauld}},
  \bibinfo{author}{\bibfnamefont{J.}~\bibnamefont{Rojo}},
  \bibinfo{author}{\bibfnamefont{L.}~\bibnamefont{Rottoli}},
  \bibinfo{author}{\bibfnamefont{S.}~\bibnamefont{Sarkar}}, \bibnamefont{and}
  \bibinfo{author}{\bibfnamefont{J.}~\bibnamefont{Talbert}},
  \bibinfo{journal}{JHEP} \textbf{\bibinfo{volume}{02}}, \bibinfo{pages}{130}
  (\bibinfo{year}{2016}), \eprint{1511.06346}.

\bibitem[{\citenamefont{Bhattacharya et~al.}(2016)\citenamefont{Bhattacharya,
  Enberg, Jeong, Kim, Reno, Sarcevic, and Stasto}}]{Bhattacharya:2016jce}
\bibinfo{author}{\bibfnamefont{A.}~\bibnamefont{Bhattacharya}},
  \bibinfo{author}{\bibfnamefont{R.}~\bibnamefont{Enberg}},
  \bibinfo{author}{\bibfnamefont{Y.~S.} \bibnamefont{Jeong}},
  \bibinfo{author}{\bibfnamefont{C.~S.} \bibnamefont{Kim}},
  \bibinfo{author}{\bibfnamefont{M.~H.} \bibnamefont{Reno}},
  \bibinfo{author}{\bibfnamefont{I.}~\bibnamefont{Sarcevic}}, \bibnamefont{and}
  \bibinfo{author}{\bibfnamefont{A.}~\bibnamefont{Stasto}}
  (\bibinfo{year}{2016}), \eprint{1607.00193}.

\bibitem[{\citenamefont{d'Enterria et~al.}(2011)\citenamefont{d'Enterria,
  Engel, Pierog, Ostapchenko, and Werner}}]{d'Enterria:2011kw}
\bibinfo{author}{\bibfnamefont{D.}~\bibnamefont{d'Enterria}},
  \bibinfo{author}{\bibfnamefont{R.}~\bibnamefont{Engel}},
  \bibinfo{author}{\bibfnamefont{T.}~\bibnamefont{Pierog}},
  \bibinfo{author}{\bibfnamefont{S.}~\bibnamefont{Ostapchenko}},
  \bibnamefont{and} \bibinfo{author}{\bibfnamefont{K.}~\bibnamefont{Werner}},
  \bibinfo{journal}{Astropart. Phys.} \textbf{\bibinfo{volume}{35}},
  \bibinfo{pages}{98} (\bibinfo{year}{2011}), \eprint{1101.5596}.

\bibitem[{\citenamefont{Abelleira~Fernandez
  et~al.}(2012)}]{AbelleiraFernandez:2012cc}
\bibinfo{author}{\bibfnamefont{J.}~\bibnamefont{Abelleira~Fernandez}}
  \bibnamefont{et~al.} (\bibinfo{collaboration}{LHeC Study Group}),
  \bibinfo{journal}{J.Phys.} \textbf{\bibinfo{volume}{G39}},
  \bibinfo{pages}{075001} (\bibinfo{year}{2012}), \eprint{1206.2913}.

\bibitem[{\citenamefont{Mangano et~al.}(2016)}]{Mangano:2016jyj}
\bibinfo{author}{\bibfnamefont{M.~L.} \bibnamefont{Mangano}}
  \bibnamefont{et~al.} (\bibinfo{year}{2016}), \eprint{1607.01831}.

\bibitem[{\citenamefont{Abramowicz et~al.}(2013)}]{Abramowicz:1900rp}
\bibinfo{author}{\bibfnamefont{H.}~\bibnamefont{Abramowicz}}
  \bibnamefont{et~al.} (\bibinfo{collaboration}{H1 , ZEUS}),
  \bibinfo{journal}{Eur.Phys.J.} \textbf{\bibinfo{volume}{C73}},
  \bibinfo{pages}{2311} (\bibinfo{year}{2013}), \eprint{1211.1182}.

\bibitem[{\citenamefont{Abramowicz et~al.}(2015)}]{Abramowicz:2015mha}
\bibinfo{author}{\bibfnamefont{H.}~\bibnamefont{Abramowicz}}
  \bibnamefont{et~al.} (\bibinfo{collaboration}{ZEUS, H1}),
  \bibinfo{journal}{Eur. Phys. J.} \textbf{\bibinfo{volume}{C75}},
  \bibinfo{pages}{580} (\bibinfo{year}{2015}), \eprint{1506.06042}.

\bibitem[{\citenamefont{Gauld et~al.}(2015)\citenamefont{Gauld, Rojo, Rottoli,
  and Talbert}}]{Gauld:2015yia}
\bibinfo{author}{\bibfnamefont{R.}~\bibnamefont{Gauld}},
  \bibinfo{author}{\bibfnamefont{J.}~\bibnamefont{Rojo}},
  \bibinfo{author}{\bibfnamefont{L.}~\bibnamefont{Rottoli}}, \bibnamefont{and}
  \bibinfo{author}{\bibfnamefont{J.}~\bibnamefont{Talbert}},
  \bibinfo{journal}{JHEP} \textbf{\bibinfo{volume}{11}}, \bibinfo{pages}{009}
  (\bibinfo{year}{2015}), \eprint{1506.08025}.

\bibitem[{\citenamefont{Zenaiev et~al.}(2015)}]{Zenaiev:2015rfa}
\bibinfo{author}{\bibfnamefont{O.}~\bibnamefont{Zenaiev}} \bibnamefont{et~al.}
  (\bibinfo{collaboration}{PROSA}), \bibinfo{journal}{Eur. Phys. J.}
  \textbf{\bibinfo{volume}{C75}}, \bibinfo{pages}{396} (\bibinfo{year}{2015}),
  \eprint{1503.04581}.

\bibitem[{\citenamefont{Cacciari et~al.}(2015)\citenamefont{Cacciari, Mangano,
  and Nason}}]{Cacciari:2015fta}
\bibinfo{author}{\bibfnamefont{M.}~\bibnamefont{Cacciari}},
  \bibinfo{author}{\bibfnamefont{M.~L.} \bibnamefont{Mangano}},
  \bibnamefont{and} \bibinfo{author}{\bibfnamefont{P.}~\bibnamefont{Nason}},
  \bibinfo{journal}{Eur. Phys. J.} \textbf{\bibinfo{volume}{C75}},
  \bibinfo{pages}{610} (\bibinfo{year}{2015}), \eprint{1507.06197}.

\bibitem[{\citenamefont{Aaij et~al.}(2013)}]{Aaij:2013mga}
\bibinfo{author}{\bibfnamefont{R.}~\bibnamefont{Aaij}} \bibnamefont{et~al.}
  (\bibinfo{collaboration}{LHCb}), \bibinfo{journal}{Nucl.Phys.}
  \textbf{\bibinfo{volume}{B871}}, \bibinfo{pages}{1} (\bibinfo{year}{2013}),
  \eprint{1302.2864}.

\bibitem[{\citenamefont{Mangano and Rojo}(2012)}]{Mangano:2012mh}
\bibinfo{author}{\bibfnamefont{M.~L.} \bibnamefont{Mangano}} \bibnamefont{and}
  \bibinfo{author}{\bibfnamefont{J.}~\bibnamefont{Rojo}},
  \bibinfo{journal}{JHEP} \textbf{\bibinfo{volume}{1208}}, \bibinfo{pages}{010}
  (\bibinfo{year}{2012}), \eprint{1206.3557}.

\bibitem[{\citenamefont{Aaij et~al.}(2016{\natexlab{a}})}]{Aaij:2016jht}
\bibinfo{author}{\bibfnamefont{R.}~\bibnamefont{Aaij}} \bibnamefont{et~al.}
  (\bibinfo{collaboration}{LHCb}) (\bibinfo{year}{2016}{\natexlab{a}}),
  \eprint{1610.02230}.

\bibitem[{\citenamefont{Aaij et~al.}(2016{\natexlab{b}})}]{Aaij:2015bpa}
\bibinfo{author}{\bibfnamefont{R.}~\bibnamefont{Aaij}} \bibnamefont{et~al.}
  (\bibinfo{collaboration}{LHCb}), \bibinfo{journal}{JHEP}
  \textbf{\bibinfo{volume}{03}}, \bibinfo{pages}{159}
  (\bibinfo{year}{2016}{\natexlab{b}}), \bibinfo{note}{[Erratum:
  JHEP05,074(2017)]}, \eprint{1510.01707}.

\bibitem[{\citenamefont{Ball et~al.}(2015)}]{Ball:2014uwa}
\bibinfo{author}{\bibfnamefont{R.~D.} \bibnamefont{Ball}} \bibnamefont{et~al.}
  (\bibinfo{collaboration}{NNPDF}), \bibinfo{journal}{JHEP}
  \textbf{\bibinfo{volume}{1504}}, \bibinfo{pages}{040} (\bibinfo{year}{2015}),
  \eprint{1410.8849}.

\bibitem[{\citenamefont{Cacciari and Greco}(1994)}]{Cacciari:1993mq}
\bibinfo{author}{\bibfnamefont{M.}~\bibnamefont{Cacciari}} \bibnamefont{and}
  \bibinfo{author}{\bibfnamefont{M.}~\bibnamefont{Greco}},
  \bibinfo{journal}{Nucl.Phys.} \textbf{\bibinfo{volume}{B421}},
  \bibinfo{pages}{530} (\bibinfo{year}{1994}), \eprint{hep-ph/9311260}.

\bibitem[{\citenamefont{Nason}(2004)}]{Nason:2004rx}
\bibinfo{author}{\bibfnamefont{P.}~\bibnamefont{Nason}},
  \bibinfo{journal}{JHEP} \textbf{\bibinfo{volume}{0411}}, \bibinfo{pages}{040}
  (\bibinfo{year}{2004}), \eprint{hep-ph/0409146}.

\bibitem[{\citenamefont{Frixione
  et~al.}(2007{\natexlab{a}})\citenamefont{Frixione, Nason, and
  Oleari}}]{Frixione:2007vw}
\bibinfo{author}{\bibfnamefont{S.}~\bibnamefont{Frixione}},
  \bibinfo{author}{\bibfnamefont{P.}~\bibnamefont{Nason}}, \bibnamefont{and}
  \bibinfo{author}{\bibfnamefont{C.}~\bibnamefont{Oleari}},
  \bibinfo{journal}{JHEP} \textbf{\bibinfo{volume}{0711}}, \bibinfo{pages}{070}
  (\bibinfo{year}{2007}{\natexlab{a}}), \eprint{0709.2092}.

\bibitem[{\citenamefont{Alioli et~al.}(2010)\citenamefont{Alioli, Nason,
  Oleari, and Re}}]{Alioli:2010xd}
\bibinfo{author}{\bibfnamefont{S.}~\bibnamefont{Alioli}},
  \bibinfo{author}{\bibfnamefont{P.}~\bibnamefont{Nason}},
  \bibinfo{author}{\bibfnamefont{C.}~\bibnamefont{Oleari}}, \bibnamefont{and}
  \bibinfo{author}{\bibfnamefont{E.}~\bibnamefont{Re}}, \bibinfo{journal}{JHEP}
  \textbf{\bibinfo{volume}{1006}}, \bibinfo{pages}{043} (\bibinfo{year}{2010}),
  \eprint{1002.2581}.

\bibitem[{\citenamefont{Frixione
  et~al.}(2007{\natexlab{b}})\citenamefont{Frixione, Nason, and
  Ridolfi}}]{Frixione:2007nw}
\bibinfo{author}{\bibfnamefont{S.}~\bibnamefont{Frixione}},
  \bibinfo{author}{\bibfnamefont{P.}~\bibnamefont{Nason}}, \bibnamefont{and}
  \bibinfo{author}{\bibfnamefont{G.}~\bibnamefont{Ridolfi}},
  \bibinfo{journal}{JHEP} \textbf{\bibinfo{volume}{0709}}, \bibinfo{pages}{126}
  (\bibinfo{year}{2007}{\natexlab{b}}), \eprint{0707.3088}.

\bibitem[{\citenamefont{Sjostrand et~al.}(2008)\citenamefont{Sjostrand, Mrenna,
  and Skands}}]{Sjostrand:2007gs}
\bibinfo{author}{\bibfnamefont{T.}~\bibnamefont{Sjostrand}},
  \bibinfo{author}{\bibfnamefont{S.}~\bibnamefont{Mrenna}}, \bibnamefont{and}
  \bibinfo{author}{\bibfnamefont{P.~Z.} \bibnamefont{Skands}},
  \bibinfo{journal}{Comput. Phys. Commun.} \textbf{\bibinfo{volume}{178}},
  \bibinfo{pages}{852} (\bibinfo{year}{2008}), \eprint{0710.3820}.

\bibitem[{\citenamefont{Sjöstrand et~al.}(2015)\citenamefont{Sjöstrand, Ask,
  Christiansen, Corke, Desai et~al.}}]{Sjostrand:2014zea}
\bibinfo{author}{\bibfnamefont{T.}~\bibnamefont{Sjöstrand}},
  \bibinfo{author}{\bibfnamefont{S.}~\bibnamefont{Ask}},
  \bibinfo{author}{\bibfnamefont{J.~R.} \bibnamefont{Christiansen}},
  \bibinfo{author}{\bibfnamefont{R.}~\bibnamefont{Corke}},
  \bibinfo{author}{\bibfnamefont{N.}~\bibnamefont{Desai}},
  \bibnamefont{et~al.}, \bibinfo{journal}{Comput.Phys.Commun.}
  \textbf{\bibinfo{volume}{191}}, \bibinfo{pages}{159} (\bibinfo{year}{2015}),
  \eprint{1410.3012}.

\bibitem[{\citenamefont{Cacciari et~al.}(2012)\citenamefont{Cacciari, Frixione,
  Houdeau, Mangano, Nason et~al.}}]{Cacciari:2012ny}
\bibinfo{author}{\bibfnamefont{M.}~\bibnamefont{Cacciari}},
  \bibinfo{author}{\bibfnamefont{S.}~\bibnamefont{Frixione}},
  \bibinfo{author}{\bibfnamefont{N.}~\bibnamefont{Houdeau}},
  \bibinfo{author}{\bibfnamefont{M.~L.} \bibnamefont{Mangano}},
  \bibinfo{author}{\bibfnamefont{P.}~\bibnamefont{Nason}},
  \bibnamefont{et~al.}, \bibinfo{journal}{JHEP}
  \textbf{\bibinfo{volume}{1210}}, \bibinfo{pages}{137} (\bibinfo{year}{2012}),
  \eprint{1205.6344}.

\bibitem[{\citenamefont{Frixione and Webber}(2002)}]{Frixione:2002ik}
\bibinfo{author}{\bibfnamefont{S.}~\bibnamefont{Frixione}} \bibnamefont{and}
  \bibinfo{author}{\bibfnamefont{B.~R.} \bibnamefont{Webber}},
  \bibinfo{journal}{JHEP} \textbf{\bibinfo{volume}{0206}}, \bibinfo{pages}{029}
  (\bibinfo{year}{2002}), \eprint{hep-ph/0204244}.

\bibitem[{\citenamefont{Alwall et~al.}(2014)\citenamefont{Alwall, Frederix,
  Frixione, Hirschi, Maltoni et~al.}}]{Alwall:2014hca}
\bibinfo{author}{\bibfnamefont{J.}~\bibnamefont{Alwall}},
  \bibinfo{author}{\bibfnamefont{R.}~\bibnamefont{Frederix}},
  \bibinfo{author}{\bibfnamefont{S.}~\bibnamefont{Frixione}},
  \bibinfo{author}{\bibfnamefont{V.}~\bibnamefont{Hirschi}},
  \bibinfo{author}{\bibfnamefont{F.}~\bibnamefont{Maltoni}},
  \bibnamefont{et~al.}, \bibinfo{journal}{JHEP}
  \textbf{\bibinfo{volume}{1407}}, \bibinfo{pages}{079} (\bibinfo{year}{2014}),
  \eprint{1405.0301}.

\bibitem[{\citenamefont{Cacciari et~al.}(1998)\citenamefont{Cacciari, Greco,
  and Nason}}]{Cacciari:1998it}
\bibinfo{author}{\bibfnamefont{M.}~\bibnamefont{Cacciari}},
  \bibinfo{author}{\bibfnamefont{M.}~\bibnamefont{Greco}}, \bibnamefont{and}
  \bibinfo{author}{\bibfnamefont{P.}~\bibnamefont{Nason}},
  \bibinfo{journal}{JHEP} \textbf{\bibinfo{volume}{9805}}, \bibinfo{pages}{007}
  (\bibinfo{year}{1998}), \eprint{hep-ph/9803400}.

\bibitem[{\citenamefont{Cacciari et~al.}(2001)\citenamefont{Cacciari, Frixione,
  and Nason}}]{Cacciari:2001td}
\bibinfo{author}{\bibfnamefont{M.}~\bibnamefont{Cacciari}},
  \bibinfo{author}{\bibfnamefont{S.}~\bibnamefont{Frixione}}, \bibnamefont{and}
  \bibinfo{author}{\bibfnamefont{P.}~\bibnamefont{Nason}},
  \bibinfo{journal}{JHEP} \textbf{\bibinfo{volume}{0103}}, \bibinfo{pages}{006}
  (\bibinfo{year}{2001}), \eprint{hep-ph/0102134}.

\bibitem[{\citenamefont{Buckley et~al.}(2015)\citenamefont{Buckley, Ferrando,
  Lloyd, Nordström, Page et~al.}}]{Buckley:2014ana}
\bibinfo{author}{\bibfnamefont{A.}~\bibnamefont{Buckley}},
  \bibinfo{author}{\bibfnamefont{J.}~\bibnamefont{Ferrando}},
  \bibinfo{author}{\bibfnamefont{S.}~\bibnamefont{Lloyd}},
  \bibinfo{author}{\bibfnamefont{K.}~\bibnamefont{Nordström}},
  \bibinfo{author}{\bibfnamefont{B.}~\bibnamefont{Page}}, \bibnamefont{et~al.},
  \bibinfo{journal}{Eur.Phys.J.} \textbf{\bibinfo{volume}{C75}},
  \bibinfo{pages}{132} (\bibinfo{year}{2015}), \eprint{1412.7420}.

\bibitem[{\citenamefont{de~Florian et~al.}(2016)}]{deFlorian:2016spz}
\bibinfo{author}{\bibfnamefont{D.}~\bibnamefont{de~Florian}}
  \bibnamefont{et~al.} (\bibinfo{collaboration}{The LHC Higgs Cross Section
  Working Group}) (\bibinfo{year}{2016}), \eprint{1610.07922}.

\bibitem[{\citenamefont{Ball et~al.}(2011)}]{Ball:2010gb}
\bibinfo{author}{\bibfnamefont{R.~D.} \bibnamefont{Ball}} \bibnamefont{et~al.}
  (\bibinfo{collaboration}{The NNPDF}), \bibinfo{journal}{Nucl. Phys.}
  \textbf{\bibinfo{volume}{B849}}, \bibinfo{pages}{112} (\bibinfo{year}{2011}),
  \eprint{1012.0836}.

\bibitem[{\citenamefont{Ball et~al.}(2012)\citenamefont{Ball, Bertone, Cerutti,
  Del~Debbio, Forte et~al.}}]{Ball:2011gg}
\bibinfo{author}{\bibfnamefont{R.~D.} \bibnamefont{Ball}},
  \bibinfo{author}{\bibfnamefont{V.}~\bibnamefont{Bertone}},
  \bibinfo{author}{\bibfnamefont{F.}~\bibnamefont{Cerutti}},
  \bibinfo{author}{\bibfnamefont{L.}~\bibnamefont{Del~Debbio}},
  \bibinfo{author}{\bibfnamefont{S.}~\bibnamefont{Forte}},
  \bibnamefont{et~al.}, \bibinfo{journal}{Nucl.Phys.}
  \textbf{\bibinfo{volume}{B855}}, \bibinfo{pages}{608} (\bibinfo{year}{2012}),
  \eprint{1108.1758}.

\bibitem[{\citenamefont{Czakon et~al.}(2016)\citenamefont{Czakon, Fiedler,
  Heymes, and Mitov}}]{Czakon:2016ckf}
\bibinfo{author}{\bibfnamefont{M.}~\bibnamefont{Czakon}},
  \bibinfo{author}{\bibfnamefont{P.}~\bibnamefont{Fiedler}},
  \bibinfo{author}{\bibfnamefont{D.}~\bibnamefont{Heymes}}, \bibnamefont{and}
  \bibinfo{author}{\bibfnamefont{A.}~\bibnamefont{Mitov}},
  \bibinfo{journal}{JHEP} \textbf{\bibinfo{volume}{05}}, \bibinfo{pages}{034}
  (\bibinfo{year}{2016}), \eprint{1601.05375}.

\bibitem[{\citenamefont{Bertone et~al.}(2014)\citenamefont{Bertone, Carrazza,
  and Rojo}}]{Bertone:2013vaa}
\bibinfo{author}{\bibfnamefont{V.}~\bibnamefont{Bertone}},
  \bibinfo{author}{\bibfnamefont{S.}~\bibnamefont{Carrazza}}, \bibnamefont{and}
  \bibinfo{author}{\bibfnamefont{J.}~\bibnamefont{Rojo}},
  \bibinfo{journal}{Comput.Phys.Commun.} \textbf{\bibinfo{volume}{185}},
  \bibinfo{pages}{1647} (\bibinfo{year}{2014}), \eprint{1310.1394}.

\bibitem[{\citenamefont{Forte et~al.}(2010)\citenamefont{Forte, Laenen, Nason,
  and Rojo}}]{Forte:2010ta}
\bibinfo{author}{\bibfnamefont{S.}~\bibnamefont{Forte}},
  \bibinfo{author}{\bibfnamefont{E.}~\bibnamefont{Laenen}},
  \bibinfo{author}{\bibfnamefont{P.}~\bibnamefont{Nason}}, \bibnamefont{and}
  \bibinfo{author}{\bibfnamefont{J.}~\bibnamefont{Rojo}},
  \bibinfo{journal}{Nucl. Phys.} \textbf{\bibinfo{volume}{B834}},
  \bibinfo{pages}{116} (\bibinfo{year}{2010}), \eprint{1001.2312}.

\bibitem[{\citenamefont{Rojo and Caola}(2009)}]{Rojo:2009ut}
\bibinfo{author}{\bibfnamefont{J.}~\bibnamefont{Rojo}} \bibnamefont{and}
  \bibinfo{author}{\bibfnamefont{F.}~\bibnamefont{Caola}}
  (\bibinfo{year}{2009}), \eprint{0906.2079}.

\bibitem[{\citenamefont{Aartsen et~al.}(2016)}]{Aartsen:2016ngq}
\bibinfo{author}{\bibfnamefont{M.~G.} \bibnamefont{Aartsen}}
  \bibnamefont{et~al.} (\bibinfo{collaboration}{IceCube})
  (\bibinfo{year}{2016}), \eprint{1607.05886}.

\bibitem[{\citenamefont{Adrian-Martinez
  et~al.}(2016)}]{Adrian-Martinez:2016fdl}
\bibinfo{author}{\bibfnamefont{S.}~\bibnamefont{Adrian-Martinez}}
  \bibnamefont{et~al.} (\bibinfo{collaboration}{KM3Net}), \bibinfo{journal}{J.
  Phys.} \textbf{\bibinfo{volume}{G43}}, \bibinfo{pages}{084001}
  (\bibinfo{year}{2016}), \eprint{1601.07459}.

\bibitem[{\citenamefont{Ball and Forte}(1994)}]{das}
\bibinfo{author}{\bibfnamefont{R.~D.} \bibnamefont{Ball}} \bibnamefont{and}
  \bibinfo{author}{\bibfnamefont{S.}~\bibnamefont{Forte}},
  \bibinfo{journal}{Phys. Lett.} \textbf{\bibinfo{volume}{B335}},
  \bibinfo{pages}{77} (\bibinfo{year}{1994}), \eprint{hep-ph/9405320}.

\bibitem[{\citenamefont{Albacete et~al.}(2015)\citenamefont{Albacete, Illana,
  and Soto-Ontoso}}]{Albacete:2015zra}
\bibinfo{author}{\bibfnamefont{J.~L.} \bibnamefont{Albacete}},
  \bibinfo{author}{\bibfnamefont{J.~I.} \bibnamefont{Illana}},
  \bibnamefont{and}
  \bibinfo{author}{\bibfnamefont{A.}~\bibnamefont{Soto-Ontoso}},
  \bibinfo{journal}{Phys. Rev.} \textbf{\bibinfo{volume}{D92}},
  \bibinfo{pages}{014027} (\bibinfo{year}{2015}), \eprint{1505.06583}.

\bibitem[{\citenamefont{Connolly et~al.}(2011)\citenamefont{Connolly, Thorne,
  and Waters}}]{Connolly:2011vc}
\bibinfo{author}{\bibfnamefont{A.}~\bibnamefont{Connolly}},
  \bibinfo{author}{\bibfnamefont{R.~S.} \bibnamefont{Thorne}},
  \bibnamefont{and} \bibinfo{author}{\bibfnamefont{D.}~\bibnamefont{Waters}},
  \bibinfo{journal}{Phys. Rev.} \textbf{\bibinfo{volume}{D83}},
  \bibinfo{pages}{113009} (\bibinfo{year}{2011}), \eprint{1102.0691}.

\bibitem[{\citenamefont{Anchordoqui et~al.}(2014)}]{Anchordoqui:2013dnh}
\bibinfo{author}{\bibfnamefont{L.~A.} \bibnamefont{Anchordoqui}}
  \bibnamefont{et~al.}, \bibinfo{journal}{JHEAp}
  \textbf{\bibinfo{volume}{1-2}}, \bibinfo{pages}{1} (\bibinfo{year}{2014}),
  \eprint{1312.6587}.

\bibitem[{\citenamefont{Jones et~al.}(2016)\citenamefont{Jones, Martin, Ryskin,
  and Teubner}}]{Jones:2016ldq}
\bibinfo{author}{\bibfnamefont{S.~P.} \bibnamefont{Jones}},
  \bibinfo{author}{\bibfnamefont{A.~D.} \bibnamefont{Martin}},
  \bibinfo{author}{\bibfnamefont{M.~G.} \bibnamefont{Ryskin}},
  \bibnamefont{and} \bibinfo{author}{\bibfnamefont{T.}~\bibnamefont{Teubner}}
  (\bibinfo{year}{2016}), \eprint{1610.02272}.

\bibitem[{\citenamefont{d'Enterria and Rojo}(2012)}]{d'Enterria:2012yj}
\bibinfo{author}{\bibfnamefont{D.}~\bibnamefont{d'Enterria}} \bibnamefont{and}
  \bibinfo{author}{\bibfnamefont{J.}~\bibnamefont{Rojo}},
  \bibinfo{journal}{Nucl.Phys.} \textbf{\bibinfo{volume}{B860}},
  \bibinfo{pages}{311} (\bibinfo{year}{2012}), \eprint{1202.1762}.

\bibitem[{\citenamefont{Peitzmann}(2016)}]{Peitzmann:2016gkt}
\bibinfo{author}{\bibfnamefont{T.}~\bibnamefont{Peitzmann}}
  (\bibinfo{collaboration}{ALICE FoCal}) (\bibinfo{year}{2016}),
  \eprint{1607.01673}.

\bibitem[{\citenamefont{Gauld}(2017)}]{Gauld:2017omh}
\bibinfo{author}{\bibfnamefont{R.}~\bibnamefont{Gauld}},
  \bibinfo{journal}{JHEP} \textbf{\bibinfo{volume}{05}}, \bibinfo{pages}{084}
  (\bibinfo{year}{2017}), \eprint{1703.03636}.

\end{thebibliography}

\end{document}